\documentclass[11pt,twoside]{article}
\usepackage{asp2014}

\aspSuppressVolSlug
\resetcounters
\newcommand{\water}{H$_2$O}
\newcommand{\methanol}{CH$_3$OH}
\newcommand{\ammonia}{NH$_3$}
\newcommand{\formaldehyde}{H$_2$CO}

\newcommand{\kms}{km~s$^{-1}$}

\begin{document}

\title{Science with an ngVLA: Understanding Massive Star Formation through Maser Imaging}

\author{Todd R. Hunter,$^1$, Crystal L. Brogan$^1$, Anna Bartkiewicz$^2$, James O. Chibueze$^{3,4,5}$,  Claudia J. Cyganowski$^6$, Tomoya Hirota$^7$, Gordon C. MacLeod$^8$, Alberto Sanna$^9$ and Jos\'e-Maria Torrelles$^{10}$\\
\affil{$^1$NRAO, Charlottesville, VA, USA; \email{thunter@nrao.edu, cbrogan@nrao.edu}}
\affil{$^2$Nicolaus Copernicus University, Torun, Poland; \email{annan@astro.umk.pl}}
\affil{$^3$Department of Physics and Astronomy, University of Nigeria, Nsukka, Nigeria}
\affil{$^4$SKA South Africa, 
Cape Town, South Africa; \email{james.chibueze@gmail.com}}
\affil{$^5$Centre for Space Research, Physics Department, North-West University, Potchefstroom, South Africa}
\affil{$^6$SUPA, School of Physics and Astronomy, University of St. Andrews, St. Andrews, UK; \email{cc243@st-andrews.ac.uk}} 
\affil{$^7$National Astronomical Observatory of Japan, Mitaka, Tokyo, Japan; \email{tomoya.hirota@nao.ac.jp}}
\affil{$^8$Hartebeesthoek Radio Astronomy Observatory, Krugersdorp,
South Africa; \email{gord@hartrao.ac.za}}
\affil{$^9$Max-Planck-Institut f\"ur Radioastronomie, Bonn, Germany; \email{asanna@mpifr-bonn.mpg.de}}
\affil{$^{10}$Institut de Ci\'encies de l'Espai,  Barcelona, Spain; \email{chema.torrelles@ice.cat}}
}

\paperauthor{Todd R. Hunter}{thunter@nrao.edu}{ 0000-0001-6492-0090O}{NRAO}{520 Edgemont Rd.}{Charlottesville}{VA}{22903}{USA}
\paperauthor{Crystal L. Brogan}{cbrogan@nrao.edu}{0000-0002-6558-7653}{NRAO}{520 Edgemont Rd.}{Charlottesville}{VA}{22903}{USA}
\paperauthor{Anna Bartkiewicz}{annan@astro.umk.pl}{}{Nicolaus Copernicus University}{}{Torun}{}{}{Poland}
\paperauthor{James O. Chibueze}{james.chibueze@gmail.com}{0000-0002-9875-7436}{SKA South Africa}{Park Road}{Pinelands}{Cape Town}{7405}{South Africa}
\paperauthor{Claudia J. Cyganowski}{cc243@st-andrews.ac.uk}{0000-0001-6725-1734}{University of St. Andrews}{}{St.Andrews}{Scotland}{KY169SS}{UK}
\paperauthor{Tomoya Hirota}{tomoya.hirota@nao.ac.jp}{}{National Astronomical Observatory of Japan}{Osawa 2-21-1}{Mitaka}{Tokyo}{181-8588}{Japan}
\paperauthor{Gordon C. MacLeod}{gord@hartrao.ac.za}{}{Hartebeesthoek Radio Astronomy Observatory}{}{Krugersdrop}{}{}{South Africa}
\paperauthor{Alberto Sanna}{asanna@mpifr-bonn.mpg.de}{0000-0001-7960-4912O}{Max-Planck-Institut f\"ur Radioastronomie}{}{Bonn}{}{53121}{Germany}
\paperauthor{Jos\'e-Maria Torrelles}{çhema.torrelles@ice.cat}{0000-0002-6896-6085}{Institut de Ciències de l’Espai}{}{Barcelona}{}{E-08193}{Spain}

\begin{abstract}
Imaging the bright maser emission produced by several molecular species at centimeter wavelengths is an essential tool for understanding the process of massive star formation because it provides a way to probe the kinematics of dense molecular gas at high angular resolution.  Unimpeded by the high dust optical depths that affect shorter wavelength observations, the high brightness temperature of these emission lines offers a way to resolve accretion and outflow motions down to scales as fine as $\sim$1-10 au in deeply embedded Galactic star-forming regions, and at sub-pc scales in nearby galaxies.  The Next Generation Very Large Array will provide the capabilities needed to fully exploit these powerful tracers.

\end{abstract}

\section{Introduction: The Scientific Importance of Masers}

The process of star formation leads to the concentration of molecular gas to high densities in molecular cloud cores.  The potential energy released by gravitational collapse and accretion onto the central protostars heats and excites the surrounding material through infrared radiation and high velocity bipolar outflows.  Both of these feedback mechanisms (radiative and mechanical) naturally produce population inversions between specific pairs of energy levels in several abundant molecules, including \water, \methanol, OH, \ammonia, SiO, and \formaldehyde.  The resulting non-thermal maser emission in the corresponding spectral transitions provides a beacon whose brightness temperature far exceeds the more commonly-excited thermal emission lines. Consequently, maser lines at centimeter wavelengths have traditionally provided a powerful probe of star formation through single-dish surveys and interferometric imaging.  In general, they trace hot, dense molecular gas, revealing the kinematics of star-forming material within a few 1000\,au from very young stars, including accretion disks and their associated jets, as well as shocks in the outflow lobes where the jets impact ambient gas. Masers are generally more prevalent in regions surrounding massive protostars, due to their higher luminosities and more energetic outflows. Furthermore, masers are
sensitive indicators of sudden changes in the pumping conditions near the protostars, and, recently, it has been recognized that maser flares, in those lines which are pumped 
by infrared photons, can be directly associated with bursts of accretion onto the stars.  In this context, maser emission provides a unique tool for probing how massive stars form, allowing us to reconstruct the gas dynamics in the vicinity of young stars with tens of Solar masses, as well as to study the accretion process in the time domain.

%

\section{Description of the problem and limits of current facilities}

With the advent of the Atacama Large Millimeter/submillimeter Array (ALMA), imaging thermal lines at high resolution has become easier,
and recent results have begun to place previous and ongoing maser studies into better physical context \citep[see, e.g., Orion Source I,][]{Plambeck16,Hirota17}.
At the distances of more typical massive star-forming regions, however, the brightness temperature sensitivity of ALMA is still not sufficient
to trace the accretion flow and accompanying jet structures that surround massive protostars, because of the high angular and spectral resolution required.
Moreover, at the shorter wavelengths of ALMA, the combination of molecular line confusion and high dust opacity toward the hot cores in protoclusters
will often block the most interesting details from ALMA's view.  In contrast, the centimeter maser transitions propagate unobscured from the innermost
regions, providing a strong signal for self-calibration, and thus enabling high dynamic range imaging on long baselines. 

\begin{figure}[ht!]   
\centering
\includegraphics[width=0.80\linewidth]{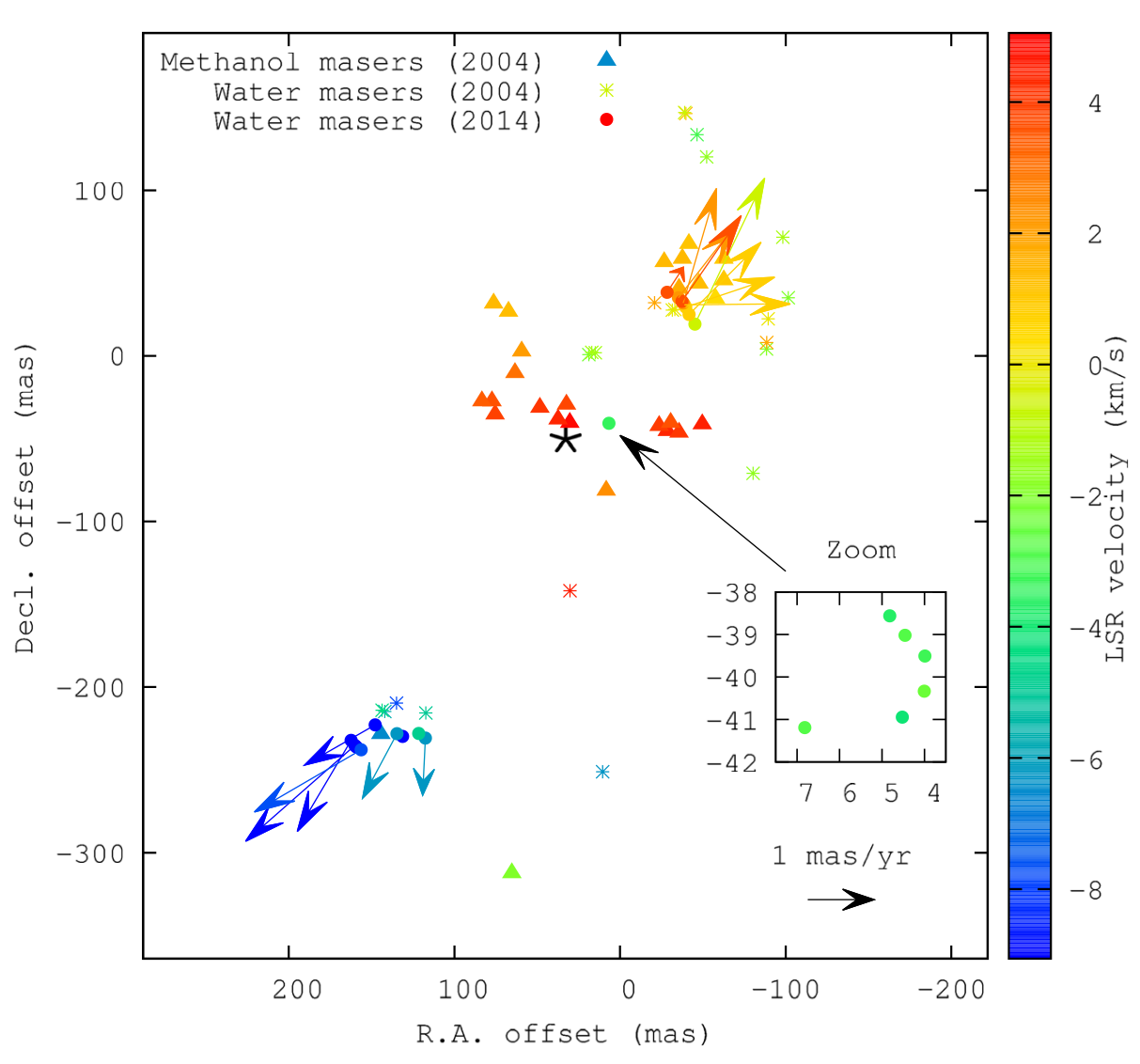}
\caption{Central region of the dominant member (MM1) of the massive protocluster AFGL~5142 (figure extracted from \citet{Burns17}). Combined view
of 22-GHz water masers (filled circles) observed with VERA in 2010, 22-GHz water masers (asterisk) observed with the VLBA in 2004 \citep{Goddi06}
and 6.7-GHz methanol masers (triangles) observed with the EVN in 2004 \citep{Goddi07}. The inset shows the trajectory of maser feature A, moving
in a clockwise fashion. The trajectory of feature A and proper motions of other masers are all converted to the YSO frame. The black asterisk symbol
indicates the approximate origin of the episodic ejections, estimated from least-squares fitting of ellipses to the VERA maser data.}
\label{afgl5142}
\end{figure}

Unfortunately, the angular resolution of the VLA is insufficient to resolve the details of accreting gas, particularly in the 6 GHz band where the beamsize
is limited to $\sim$0.3 arcsec. In the handful of nearest examples of massive star formation (d$\sim$1~kpc), this resolution corresponds to 300~au
\citep[e.g.,][]{Brogan2016}. However, in the majority of star-forming sites across the Milky Way at several kpc from the Sun, it exceeds 1000~au,
which is often more than the separation of protostars in the centers of massive protoclusters \citep[e.g.,][]{Palau13}. Thus, an order of magnitude
improvement in angular resolution (requiring at least $\sim$300~km baselines) is needed to resolve the spatial morphology and kinematics
of disks, or other accretion structures, at scales of 10-100 au in a large sample of massive protostars. Such a resolution would also enable three dimensional
measurements of gas velocity via multi-epoch proper motions.
For instance, with an angular resolution
of 10~mas in the bright H$_2$O maser line, it is possible to determine the maser positions with an accuracy better than 0.1~mas (assuming S/N $> 100$),
and then to measure proper motions of order 10~\kms\/ in a few months only, for sources located up to several kpc distance.  Furthermore, proper motion measurements of different maser species toward the same region have the potential to trace simultaneously the complementary kinematic structures around a young star, providing a unique picture of the gas dynamics locally  \citep{Sanna2010,Goddi2011}.
These measurements are currently conducted by means of Very Long Baseline Interferometry 
(VLBI) observations, but with the small number of available antennas their sensitivity is limited to non-thermal processes exceeding brightness temperatures of $\sim10^7$\,K \citep[e.g.][]{Matsumoto14,Bartkiewicz09}.  An example of proper motion
observations at these scales, using three different VLBI facilities, is shown in Fig.~\ref{afgl5142}  \citep{Burns17}.

While current VLBI facilities (VLBA, EVN, eMERLIN, KVN, VERA, and LBA) have the requisite angular resolution to trace maser proper motions accurately,
studies at these scales currently suffer from poor surface brightness sensitivity, which affects the science in two key ways.  First, only the brightest maser
spots can be detected, reducing the fidelity with which kinematic structures can be delineated in a single epoch, and reducing the number of potential
spots that will persist over multiple epochs (used for proper motion studies).  Second, the thermal radio continuum emission which arises in the immediate vicinity of  massive protostars, with flux densities of $<1$\,mJy typically \citep{Cyganowski2011}, cannot be observed with the masers simultaneously, leading to (relative) positional uncertainties between the protostellar and maser components.  The resulting ambiguity of the dynamical center severely hinders the interpretation of multi-epoch proper motion measurements, which are essential to understand the mass, momentum, and kinetic energy of the inner jet where it transitions into a bipolar molecular outflow.  

Studying additional objects at scales of 10-100~au in a comprehensive list of maser lines, and with sufficient brightness sensitivity to image simultaneously
the associated thermal free-free continuum emission, will be an important task for the next generation Very Large Array (ngVLA); see the
accompanying chapter on Jets from YSOs \citep{GalvanMadrid18}. These studies will test and expand our current picture of massive star formation into
the broader context of the Milky Way.  Moreover, with 300 km baselines, it will also be possible to trace molecular gas structures down to 0.2-pc scales
in the star-forming clouds in nearby extragalactic nuclei (e.g., Maffei~2 / IC~342 / M82) in the bright water maser and Class I methanol maser lines,
enabling the study of individual massive protoclusters in these objects for the first time. 


\section{Astronomical Impact}
	\label{impact}
While each maser transition offers a unique view into particular phenomena of massive star formation, masers can be broadly classified into
major categories. The Class II \methanol\/ maser lines, primarily at 6.7~GHz, 12.2~GHz, and 19.9~GHz, trace hot molecular gas that is (at least) moderately close
to the youngest massive protostars ($\lesssim$ 1000~au), such that they can provide sufficiently intense mid-infrared emission to
pump the maser transitions \citep[e.g.,][]{Moscadelli2011,Bartkiewicz14}. The light curves of this maser species also show intriguing properties. Quasi-periodic flares in one or more Class II \methanol\/ maser lines (120-500 days) have been observed
in about a dozen objects \citep[e.g.,][]{Goedhart2014}; in one case, the 4.83~GHz \formaldehyde\/ maser also shows correlated flaring
\citep{Araya2010}. Recently, two spectacular accretion outbursts in massive protostars have been accompanied by strong flaring of these
lines, S~255~NIRS3 \citep{Caratti17,Moscadelli2017} and NGC~6334I-MM1 \citep{Hunter17,Hunter18,MacLeod18,Brogan18a}, supporting the idea that maser flares might be caused by a variable accretion rate onto the central protostar. These extraordinary events led
to the formation of the international Maser Monitoring Organization (M2O), with the goal of detecting and reporting future maser flares so that interferometers can be alerted to study the accretion event while it is still underway.  Such an accretion event is also expected to yield variation in the continuum emission from the thermal jet \citep{Cesaroni18} and/or the hypercompact HII region \citep{Brogan18b}.  Since both of these phenomena are powered by the protostar, the ability of ngVLA to perform simultaneous observations of the continuum and the masers will enable direct measurements of the correlations between them, yielding important constraints on the physics of the accretion mechanism. 

The Class II \methanol\/ maser lines \citep{Sobolev97,Cragg05}, along with the 1.6~GHz ground state OH lines and several excited state OH lines which are radiatively pumped (at 4.66~GHz, 4.75~GHz, 4.765~GHz, 6.030~GHz, and 6.035~GHz), are also seen to trace the ionization front of ultracompact HII regions \citep[e.g.,][]{Fish07}, which are powered by the more evolved massive protostars and Zero-Age Main Sequence (ZAMS) OB stars.  Although excited OH lines are generally considered rare, a recent unbiased survey found that the 6.035~GHz line is detected toward nearly 30\% of Class~II \methanol\/ masers and with a similar distribution in Galactic latitude \citep{Avison16}.  A similar detection rate is seen in survey of the 4.765~GHz line \citep{Dodson02}.  A simple explanation is these excited OH masers always occur in the same objects that power Class~II masers, but simply have a correspondingly shorter mean lifetime or duty cycle, perhaps reflecting how long they remain above current sensitivity levels following each successive accretion outburst.  In rare cases, the main line OH masers can also trace outflow motion \citep[e.g., W75N and W3OH-TW][]{Fish11,Argon03}.

The strong water maser line at 22~GHz also traces gas close to massive and intermediate mass protostars. Often these lines span a broad  
(LSR) velocity range, of several tens of \kms, about the systemic velocity of the young stars, particularly compared to both classes of methanol
masers ($\lesssim10$\,\kms). In some cases, water masers clearly arise from gas in the first few hundred au of the jet, such as in
Cepheus~A \citep[e.g.,][]{Torrelles2011,Chibueze12}, or in bow shocks somewhat further out \citep[e.g.,][]{Sanna2012,Burns16}. 
With continent-scale baselines, proper motion studies of these masers reveal the 3D velocities and orientations of collimated jets and/or wide-angle 
winds in the inner few 1000\,au from the central protostars \citep[e.g.,][]{Torrelles01,Torrelles2003,Torrelles2014,Moscadelli2007,Sanna2010,Burns17}.
When these studies are combined with high-resolution radio continuum observations of radio thermal jets, they can allow us to quantify the outflow energetics
directly produced by the star formation process \citep[e.g.,][]{Moscadelli2016,Sanna2016}, as opposed to estimates of the molecular outflow energetics
that are attainable on scales greater than 0.1\,pc (typically through CO isotopologues). Long-term monitoring studies demonstrate that
water masers are also highly variable \citep[e.g.,][]{Felli07}, and their primary pumping mechanism is not believed to be radiative but 
collisional. Thus, since water masers are fundamentally produced in specific ranges of gas density and temperature within shocked gas layers,
they are likely to trace different types of coherent motions at different stages of protostellar evolution. This is the case, for instance, of the star-forming 
region W75\,N, where the 22~GHz masers (and the radio continuum) show different spatial distributions around two distinct young stars at different
evolutionary stages \citep{Torrelles2003,Carrasco2015}. The 22~GHz line is also unique in exhibiting the `superburst' phenomenon, in which brief
flares reach 10$^5$~Jy or more.  It has happened in only a few objects including Orion~KL \citep[][and references therein]{Hirota14} and
G25.65+1.05 \citep{Lekht18}, but it has repeated in both, and appears to be due to interaction of the jet with high density clumps in the ambient
gas, but within a few thousand au of the central protostar. In addition to Galactic studies, the detection and imaging of water masers in nearby
star-forming galaxies provides a powerful probe of optically-obscured areas of star formation like the overlap region of the Antennae \citep{Brogan10}.
Additional uses of extragalactic masers are described in the accompanying chapter on Megamaser Cosmology \citep{Braatz18}.

In contrast to \water\/ masers and Class~II \methanol\/ masers, the Class~I \methanol\/ maser lines (primarily at 25~GHz, 36~GHz, 37~GHz, 44~GHz, and 95~GHz, but
see \citet{Muller04} for a more complete list) typically arise from collisionally-excited gas located much further from the protostar ($\sim$0.1-0.5~pc) where the bipolar molecular outflow lobes impact ambient gas.  VLA surveys of these masers show that they often coincide with 4.5 micron emission that traces shocked gas in active outflows \citep{Towner17,Cyganowski2012,Cyganowski2009}.  
A similar maser phenomenon occurs in the ortho-\ammonia\/ (3,3) and (6,6) lines at 23.87~GHz and 25.06 GHz  \citep{Brogan11,Beuther07,Zhang99,Kraemer95}.
Unfortunately, the VLA beam size is insufficient to resolve the structure of individual Class~I maser features but VLBI observations resolve out most of the emission.
Furthermore, the sensitivity of VLBI arrays is insufficient to map weak maser features detected by the VLA and single dishes \citep[e.g.][]{Matsumoto14}.    We note that a portion of Class~II maser emission also tends to be resolved out on VLBI scales \citep{Pandian11,Bartkiewicz09,Minier02}.    Thus, the ngVLA will provide a unique tool to study the spatial and velocity structures of both types of methanol masers.

The Class~I \methanol\/ masers are particularly abundant in the Galactic Center star-forming
clouds \citep{McEwen16}, and have recently been detected in starburst galaxies \citep{Ellingsen17,McCarthy17}.  With ngVLA sensitivity and resolution,
these masers could be used to probe such star-forming clouds in nearby galaxies (including Maffei~2 / IC~342 / M82).  

Many maser lines are significantly
polarized, thus it is important to observe them with full Stokes products in order to obtain the highest fidelity imaging. Furthermore, recent full polarization
images of various Class~I and Class~II methanol maser lines \citep[e.g.,][]{Surcis2013,DallOlio17,Momjian2017} and the 22~GHz water line
\citep[e.g.,][]{Surcis2011,Goddi2017} have been used successfully to measure the magnetic field toward massive protostars via Zeeman
splitting. These maser lines thus offer the potential to help us understand the degree to which magnetic fields influence or control accretion and outflows in massive
protostars by providing very valuable information of magnetic field and velocity vectors in the same maser features \citep{Sanna2015,Goddi2017}. 

Finally, maser emission from the vibrationally-excited levels of SiO offers a powerful (though rare) probe of the innermost hot gas surrounding massive protostars.  For example, in one spectacular nearby case (Orion~KL), movies of the vibrationally-excited SiO J=1-0, v=0 and v=1 transitions at 43~GHz have revealed a complicated structure of disk rotation and outflow \citep{Matthews2010}.  Additional massive protostars (at greater distances) have recently been detected in these lines \citep{Cordiner16,Ginsburg15,Zapata09}.  The increased sensitivity of the ngVLA will no doubt yield further detections and enable new detailed images of the inner accretion structures in these objects.  Additional uses of all maser species is described in the accompanying chapter on Evolved Stars \citep{Matthews18}.

\section{Connection to Unique ngVLA Capabilities}
	
With its proposed frequency span, the ngVLA will uniquely provide access to all of the most important maser transitions from OH at 1.6~GHz to \methanol\/ at 109~GHz.  While the Long Baseline Observatory's Very Long Baseline Array (VLBA) also covers most of this frequency span, it lacks the sensitivity at the critical baseline lengths of up to 300~km to image simultaneously the continuum emission.  The ngVLA will provide the required balance between angular resolution and brightness sensitivity, filling an important gap in existing capabilities. 
With the ability to image the non-thermal and thermal processes simultaneously, it will finally be possible to link the studies of the 3D gas dynamics (using the masers) with studies of the physical conditions of the ionized and molecular gas (using the continuum and strong, compact thermal lines like ammonia, respectively) at the same spatial resolution.  Also, the ability to acquire high-fidelity images of all of these maser lines in just a few tunings will promote more uniform surveys of massive protostars as well as enable rapid monitoring of protostars currently undergoing an accretion outburst. Furthermore, the broader bandwidth receivers will provide more robust measurements of the spectral index of the continuum emission by promoting the ability to obtain all the necessary observations at a common epoch. Finally, the high spectral resolution and full polarization capability of the ngVLA will allow measurements of the magnetic field in the masing molecular gas via the Zeeman effect in methanol and water, which is a fundamental quantity for understanding the physics of star formation.

\section{Experimental layout}

In addition to studies of individual high mass protostellar objects (HMPOs), such as those currently undergoing an outburst, one can foresee a large ngVLA project to perform multi-epoch imaging of a significant sample of (several dozen) high-mass star-forming regions (HMSFRs). These HMSFRs typically contain clusters of massive protostars in diverse evolutionary states, a phenomenon that is often termed a ``proto-Trapezium'' \citep{Megeath05} or a ``protocluster'' \citep{Minier05}. The following project would allow us: (1) to take a census of the young stellar objects across a broad mass range in these protoclusters, directly measuring the initial-mass-function; (2) to measure the kinematics of the gas undergoing maser emission, which will trace the flow of gas near the most massive stars; and (3) to study the way massive stars feed energy back into the protocluster environment and influence the formation of Solar-mass stars (and vice versa), by combining points (1) and (2).

Most protoclusters will be easily encompassed by a single pointing of the ngVLA in the lower bands, in which the primary beam full width half power (FWHP) is envisioned to be $\sim$$10'$ (15~pc at 5~kpc) at 6.7~GHz.  The largest ones may require a few pointings at the highest band where the beam is only $\sim$$40''$ (1~pc at 5~kpc) at 100~GHz. Given that protostars at different stages of evolution may excite different masers, imaging these fields in all the ngVLA bands, and using all available antennas with full polarization products will efficiently identify all the massive protostars via their continuum emission (ionized gas and/or dust) and various maser lines.  To illustrate the level of sensitivity required, we consider the continuum emission from jets which are closely linked to maser activity (see \S\ref{impact}).  Furthermore, we consider specifically jets from intermediate-mass protostars in order to define a sensitivity that will probe a broad range of massive protostars.  The emission from individual clumps in jets observed by the VLA toward intermediate-mass protostars is approximately 1~mJy at 2-20~GHz in nearby (400~pc) regions like Serpens \citep{RK2016}, which translates to only 6~$\mu$Jy for similar examples that are expected to populate more massive protoclusters at 5~kpc. In order to measure the SED of such an object, we require a $\approx6\sigma$ detection in each 1~GHz subband.  With the current VLA, we would need 44~hr of VLA observing time to reach $1~\mu$Jy rms in X-band alone.  With the increased effective collecting area of the ngVLA, this time requirement would drop to $\lesssim1$~hr per band, meaning that a multi-band survey of many fields would become feasible.  Also, with this sensitivity, young lower-mass T~Tauri stars, which are chromospherically active stars associated with highly-variable faint (synchrotron) radio continuum emission, will be also detected in the lower frequency bands (e.g., C-band), providing information about the low-mass population \citep{Forbrich17}. 

The available correlator channels will be distributed accordingly among low resolution continuum windows and up to 20 high-resolution spectral line windows in each of the 6 bands.  Due to the 300~km baselines, we note that the maximum channel width for a 24~GHz (K-band) continuum observation is 0.5~MHz in order to limit bandwidth smearing to less than 2\% across the FWHP of the primary beam, which means that $\sim$28000 dual-polarization channels will be needed for the 14~GHz instantaneous bandwidth available in that band. In addition, the high-resolution line windows will require between 256 channels and 1024 channels each in order to provide Hanning-smoothed spectra with 0.1~\kms\/ resolution across a velocity width of 12~\kms\/ to 50~\kms. The broadest lines (primarily water) will require 4096 channels to cover $\sim$200~\kms, so an additional $\sim$20000 full-polarization channels are also needed for the maser lines.  Thus, the number of correlator resources needed for the line observations is comparable to that needed for the continuum. A list of maser lines detectable in each band is summarized in Table~\ref{lines}.

\begin{table}[!ht]
\caption{Detected maser lines in each ngVLA band; see Table~1 of \citet{Menten2007} for further details}
\label{lines}
\begin{center}
{\small
\begin{tabular}{cccc}
\tableline
\noalign{\smallskip}
Band & Range (GHz) & Species and line frequencies (GHz) & \# lines\\
\tableline
\noalign{\smallskip}
1 & 1.2 - 3.5   & ground-state OH (1.612, 1.665, 1.667, 1.720) & 4 \\
\tableline
2 & 3.5 - 12.3  & excited OH (4.66, 4.75, 4.765, 6.031, 6.035, 6.049) & 6 \\
  &             & \methanol\ Class~I (9.936) & 1\\
  &             & \methanol\ Class~II (6.668, 12.18) & 2\\
  &             & ortho-\formaldehyde\ (4.83) & 1 \\
  \tableline
3 & 12.3 - 20.5 & \methanol\ Class~II (19.97) & 1 \\ 
  &             & excited OH  \citep[13.441,][]{Baudry81,Caswell04} & 1 \\
\tableline
4 & 20.5 - 34   & \water\ (22.235) & 1\\
  &             & \methanol\  $J_2$-$J_1$-series Class~I  (24.9-30.3) & 15\\
     &             & \methanol\ Class~I \citep[23.445,][]{Voronkov11} & 1\\ 
   &             & \methanol\ Class~II \citep[23.12,][]{Cragg04}) & 1\\ 
  &             & ortho-\ammonia\ (3,3) (6,6) (9,9) (12,12) & 4\\
   &            & other \ammonia\ inversion lines (thermal and/or maser) & $\sim$dozen\\
    &           & excited OH  \citep[23.8,][]{Baudry81} & 1 \\
\tableline
5 & 30.5 - 50.5 & Class I \methanol\ (36.169, 44.069) & 2\\
   &              & Class II \methanol\ (37.7, 38.29, 38.45) & 3\\
  &             & SiO 1-0 v=1,2 (43.12, 42.82) & 2 \\
\tableline
6 & 70 - 116    & \methanol\ Class~I (84.521, 95.169, 104.3) & 3 \\
  &             & \methanol\ Class~II (76.2, 76.5, 85.5, 86.6, 104.1, 107, 108.8) & 7 \\
  &             & SiO 2-1 v=1,2 (86.24, 85.64) & 2\\
\tableline\
\end{tabular}
}
\end{center}
\end{table}

With the calibrated uv-data, we will first construct multi-band continuum images with a matched resolution of $0.05''$ in the range 5-100 GHz, using all the baselines ($\sim$360~km) at 5~GHz, and tapering the data to 60~km at 30~GHz and 18~km at 100~GHz.  The photometry from these images will provide the spectral energy distributions of all the individual protostars (or compact binaries) on scales of 250~au $\times (d/5~{\rm kpc})$, giving an immediate census of the protocluster.  At these scales, images of thermal line emission from tracers of warm dense gas can also be attempted.  Finer details of the individual objects can then be pursued by imaging the continuum using the longer baselines in the higher frequency bands, for instance, as fine as $0.01''$ at 22~GHz (50~au $\times$(d/5{\rm kpc})).  This step will allow us to distinguish between jets (linearly-shaped emission), ultracompact and hypercompact HII regions (more symmetric emission), and T~Tauri stars (non-thermal point-like emission).   At this scale, in the higher frequency bands, disk-like accretion structures may be visible in the (still) optically-thin dust emission. Next, we can image the full field in the individual maser lines at moderate spectral resolution ($\sim$1~\kms) and in both polarizations, tabulating which transitions are associated with each object.  In some cases, this process will include identifying the angular scale at which the emission becomes resolved in order to maintain sensitivity to the weakest features.  We can then make spatially smaller cubes of each maser line, imaging several smaller sub-fields (i.e., those containing all the strong sources) taking advantage of the highest angular and spectral resolution ($\sim$0.1~\kms) in order to analyze the maser gas kinematics associated with each protostar.  In many cases, we would make these cubes with full polarization in order to measure the magnetic fields.  Finally, the strongest compact features can then be used to measure proper motions over several months or years.  

On the one hand, such an ambitious project would require the efforts of many students and researchers at many international institutions, but on the other hand, it will foster new synergies between astronomers with diverse expertise, and provide a global picture of the youngest star-forming sites across the Milky Way.

\section{Complementarity with respect to existing and planned (>2025) facilities}
	
The science described here will share many synergies with other future facilities.  Of course, ALMA will likely be the closest complementary facility as it can observe the dust emission and strong thermal lines at comparable angular resolutions.  The Square Kilometer Array (SKA) will provide access to high resolution imaging of the lowest frequency masers, such as OH at 1612~MHz, 1665~MHz, 1667~MHz and 1720~MHz, as well as covering the southern Galactic plane in the additional maser lines that will be accessible in its highest frequency bands including Class~II methanol masers.  Toward many target regions, imaging at mid-infrared wavelengths by the Thirty Meter Telescope (TMT) and the European Extremely Large Telescope (ELT) will be quite informative as these facilities can potentially deliver 0.05-0.2 arcsec resolution at 7-28~\micron\/ which can probe through a portion of the typically high column of extinction. The James Webb Space Telescope (JWST) can also provide sensitive integral field images for regions where saturation is not prohibitive. In both cases, the short wavelengths will delineate outflow cavities on larger scales, while the longer wavelengths (when combined with ALMA observations) will enable measurements of the luminosities of the individual massive protostars. Finally, future X-ray space telescopes such as Lynx \citep{Vikhlinin18} can potentially provide crucial measurements of transient hot gas following accretion outbursts in massive protostars, as well as study the variable emission from the lower mass members of the protocluster \citep[see, e.g.,][]{Montes15,Townsley14}.

\acknowledgements The National Radio Astronomy Observatory is a facility of the National Science Foundation operated under agreement by Associated Universities, Inc.  




\end{document}